\documentclass{mem}
\usepackage{natbib}\usepackage{txfonts}\usepackage{balance}
\usepackage{graphicx}
\usepackage[a4paper]{hyperref}
\idline{75}{282}
\begin{document}
\def\teff{$T\rm_{eff }$}
\def\kms{$\mathrm {km s}^{-1}$}

\title{
RR Lyrae variables in GGCs:
distribution of periods and synthetic models.
}

   \subtitle{}

\author{
M., Castellani\inst{1}, V. Castellani\inst{1},
S. Cassisi\inst{2} and F. Caputo\inst{1}
          }

  \offprints{M. Castellani}

\institute{
INAF --
Osservatorio Astronomico di Roma, Via Frascati 33,
00040 Monteporzio Catone, Rome, Italy
  \email{m.castellani@mporzio.astro.it}
\and
INAF --
Osservatorio Astronomico di Collurania, 
via M. Maggini, 64100 Teramo, Italy
}
\authorrunning{Castellani}

\titlerunning{RRLyrae variables in GGCs}

\abstract{
We present some applications of our 
Synthetic Horizontal Branches (SHB) simulations, aimed
to reproduce the peculiar period
distributions of RR Lyrae belonging to the 
Galactic Globular Clusters  M3 and M5.
We show some evidence, supporting
the importance of SHBs
in obtaining parameters such as the mass distribution inside the
instability strip.

}
\maketitle{}

\section{Introduction}

Since Oosterhoff (1939), pulsation periods of RR Lyrae variables
in Galactic Globular Clusters (GGCs) have been the crossroad of several empirical and 
theoretical investigations (e.g. Rood 1973; Catelan 2004).
Periods 
are robust observables to constrain 
evolutionary predictions. We know that the pulsation is governed 
by the physical properties of stellar structures, thus 
providing independent 
access to the evolutionary features of low-mass stars.


On this basis, we attempted a direct connection 
between pulsation and evolutionary theories, by investigating 
the behavior of up-to-date low-mass stellar models 
evolving through the central He burning, Horizontal Branch (HB),
evolutionary phase. We accomplished this goal by adopting  
our synthetic HB (SHB) procedure (see Castellani et al. 2005
and references therein).

\section{Selected clusters: M3 \& M5}

In order
to investigate whether
canonical HB models and pulsational predictions
(periods and instability boundaries, see 
Di Criscienzo et al. 2004)
do account for the 
observed peaked distribution of RR Lyrae periods in M3,
we performed a detailed set of numerical experiments.

At variance with previous findings,
we found that by assuming a suitable bimodal mass distribution, "canonical"
models can still provide a plausible explanation 
of the observed period distribution 
for the RR Lyrae stars in M3. 
In particular, our best SHB fit model attain 
a Kolmogorov-Smirnov (KS) similarity of 99.9\%.
Such a model relies on a gaussian mass distribution 
centered on M=0.68 ${M_\odot}$
with a dispersion as narrow as $\sigma\sim$0.005, plus a flat mass
distribution of about 200 stars, distributed over the range of
masses 0.65 to 0.61 ${M_\odot}$. This component 
gives a marginal
contribution in terms of variable stars
($\leq10\%$), but it is 
required to mimic the observed rich population of blue HB stars present in
M3.


\begin{figure}[]
\resizebox{\hsize}{!}{\includegraphics[clip=true]{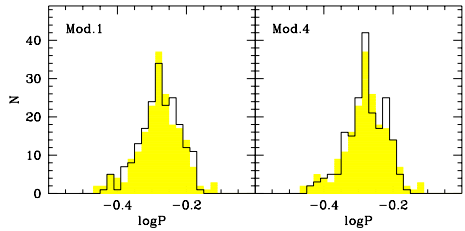}}
\caption{
\footnotesize
The comparison between our two best  
simulations and the observed period 
distribution (shaded histogram)
of M3 variables.
}
\end{figure}


However, one finds that canonical models outnumber
the observed number of red HB stars ($\sim$110, see Catelan 2004) by roughly
a factor of two. On the other hand, the 
predicted ratio between the number of RR Lyrae stars
and red HB stars depends on several 
astrophysical parameters, such as
the shape of the mass distribution 
and/or the temperature range covered by evolutionary 
tracks. In particular, by adopting a truncated gaussian (i.e., by 
neglecting  all HB masses larger than the mean mass) 
we obtain a number of red stars similar to the observed ones, 
while keeping a KS similarity of 82\% .

Interestingly, it appears that different mass distributions are needed to mimic
the period distributions of variables in different clusters.
Preliminary numerical experiments support the evidence 
that a good fit of the period distribution in M5 might be 
obtained with a “flat” mass distribution on the HB, 
and stellar masses ranging from 0.61 to 0.70 ${M_\odot}$ (see Fig. 2). 


\begin{figure}[]
\resizebox{\hsize}{!}{\includegraphics[clip=true]{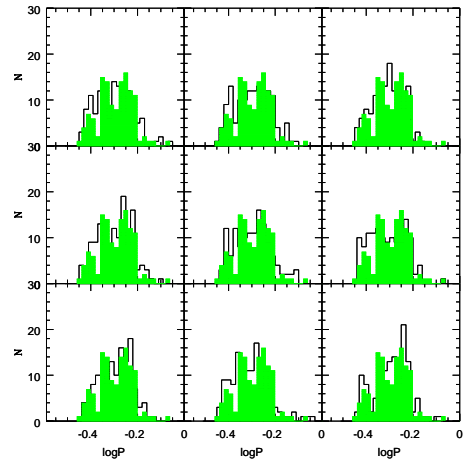}}
\caption{
\footnotesize
Same as Fig. 1, but for nine simulations of M5 period distribution.
}
\end{figure}


\section{Conclusions}
SHB appears to be a very powerful approach 
to compare predicted 
and empirical period-distribution
of cluster RR Lyrae stars.
Morover, and even more importantly, this approach 
provides the unique opportunity to constrain the 
mass distribution inside the
instability strip.
Current simulations support the evidence that a gaussian 
mass distribution provides a plausible explanation 
for the period distribution
in several GGCs. However, preliminary 
results indicate that the period distribution of 
M5 RR Lyrae stars might be explained by
adopting a flat mass distribution.

We also plan to apply the same approach to GGCs which host 
a sizable sample of RR Lyrae stars, such as Omega Cen.

\begin{acknowledgements}
We acknowledge G. Bono for his suggestions that helped
us to clearly improve the presentation of the results
of our work.
\end{acknowledgements}

\bibliographystyle{aa}

\end{document}